\documentclass[aps,pra,twocolumn,superscriptaddress]{revtex4-1}
\usepackage{subfigure}
\usepackage{xcolor}
\usepackage{graphicx}% Include figure files
\usepackage{bm}% bold mathd
\usepackage{amsmath}
\usepackage{CJK}
\usepackage{float}
\usepackage[utf8]{inputenc}

\begin{document}
\begin{CJK*}{UTF8}{gbsn}
\title{Dynamics of  momentum distribution and structure factor in a weakly interacting Bose gas with a periodic modulation}
\author{Ning Liu (刘宁)}
\thanks{Corresponding author}
\email{ningliu@mail.bnu.edu.cn}
\affiliation{Department of Physics, Beijing Normal University, Beijing 100875, China}

\author{Z. C. Tu (涂展春)}

\affiliation{Department of Physics, Beijing Normal University, Beijing 100875, China}

\date{\today}% It is always \today, today,
\begin{abstract}
 The momentum distribution and dynamical structure factor in a weakly interacting Bose gas with a time-dependent periodic modulation in terms of the Bogoliubov treatment are investigated. The evolution equation related to the Bogoliubov weights happens to be a solvable Mathieu equation when the coupling strength is periodically modulated. An exact relation between the time derivatives of momentum distribution and dynamical structure factor is derived, which indicates that the single-particle property strongly related to the two-body property in the evolutions of Bose-Einstein condensates. It is found that the momentum distribution and dynamical structure factor cannot display periodical behavior. For stable dynamics, some particular peaks in the curves of momentum distribution and dynamical structure factor appear synchronously, which is consistent with the derivative relation.
\end{abstract}

\maketitle
\end{CJK*}
\section{Introduction}
Ultracold atom gases provide pure and precise platforms to study the physics of many-body systems. The theoretical and experimental advances of ultracold atom gases in a driven potential have unprecedentedly red influenced the understanding of dynamical behaviors of many-body systems~\cite{Bloch2008, Eckardt2017}. Recently, Atas \textit{et al.} have investigated the dynamics of Tonks-Girardeau gas in a harmonic potential with a time-dependent frequency. They have proved that the solution to this problem can be mapped into the Mathieu equation~\cite{Atas2019}. As is well-known, the Tonks-Girardeau gas can not emerge Bose-Einstein condensation due to the strong interaction of the one-dimensional Bosons. For the weak-interacting Bose gas, Arnal \textit{et al.} have analyzed the micromotion of one-dimensional Bose-Einstein condensates~(BEC) in a periodically driven potential~\cite{Arnal2020}. Bala\v{z} and Nicolin have found Faraday waves in binary immiscible BEC and inhomogeneous BEC in a periodically driven radial potential, respectively~\cite{Balaz2012, Balaz2014, Sudharsan2016}.

Ultracold atom gases with a time-dependent interaction have also attracted much attention from physicists since the Feshbach resonance enables us to arbitrarily tune the magnitude of the interaction between atoms even the sign of scattering length which describes the property of interaction~\cite{Inouye1998}. Vidanovi\'{c} \textit{et al.} investigated the nonlinear dynamics of BEC induced by a harmonic modulation of the interaction, and found resonant effects in collective oscillation modes by numerical simulation of the Gross-Pitaevskii equation~\cite{Vidanovic2011}. Recently, collective emission of matter-wave jets resembling fireworks has been observed in BEC with periodically modulated coupling strength~\cite{Clark2017, Fu2018}, which was excellently explained in theory by Wu and Zhai~\cite{Wu2019}. The above significant experimental observation and its theoretical explanation stimulate us to study dynamics in an interacting Bose gas with a periodic modulation. In this paper, we employ the Bogoliubov treatment developed by Martone \textit{et al.}~\cite{Martone2018} to analytically investigate the dynamics of the weakly interacting Bose gas with a time-periodic coupling. We find that the evolution equation related to Bogoliubov weights happens to be a solvable Mathieu equation when the coupling strength is periodically modulated. We derive analytical time-evolution expressions of the momentum distribution and dynamical structure factor. Specifically, we find there is a specific equation between the derivatives of these two quantities. We demonstrate the evolution of the two quantities cannot display periodical behavior. We also show stable and unstable dynamics when the characteristic relations of the Mathieu equation are broken.

The rest of this paper is organized as follows. In section \ref{Bogo}, we briefly introduce the Bogoliubov treatment and the time-dependent harmonic oscillator method which is used to calculate the time-propagated Bogoliubov weights. In section \ref{Peri}, we focus on the dynamical equation of the evolution function related to Bogoliubov weights, which is the key equation of the whole theory. The equation of evolution function with a cosine-varied coupling constant is found to be the Mathieu equation. We derive the expressions of momentum distribution and dynamical structure factor and an equation between their derivatives. And then, we show the different kinds of time evolutions of momentum distribution and dynamical structure factor in section \ref{ens}, ie., the stable and unstable dynamics with the choices of particular experimentally-modulated parameters. A summary is given in section \ref{conc}.
%%%%========================================================================
\section{Bogoliubov treatment for evolution equation}\label{Bogo}
%%%%========================================================================
Martone \textit{et~al.} developed the Bogoliubov treatment and then calculated three exactly solvable models including steplike coupling, Woods-Saxon coupling, and modified P\"{o}schl-Teller coupling. In this section, we follow~\cite{Martone2018} and briefly introduce the Bogoliubov treatment and time-dependent harmonic oscillator method.

For a weakly interacting uniform Bose gas, after Bogoliubov approximation, the Hamilton is
\begin{equation}
\begin{aligned}
  H&=E_0+\sum_{\bm{k}\ne 0}(\epsilon_{{k}}+g(t)\rho)a_{\bm{k}}^\dagger a_{\bm{k}}\\
  &+\frac{g(t)\rho}{2}\sum_{\bm{k}\ne 0}(a_{\bm{k}}^\dagger a_{-\bm{k}}^\dagger+a_{\bm{k}}a_{-\bm{k}}
  ),\label{H}
\end{aligned}
\end{equation}
where $E_0$ is mean-field ground energy, $\epsilon_{{k}}=\hbar^2k^2/2m$. $g(t)$ represents time-dependent coupling strength, which is proportional to scatter length. Then we can obtain Heisenberg equation of $a_{\bm{k}},a_{-\bm{k}}^\dagger$ from $H$,
\begin{equation}
  {\rm i}\hbar\frac{{\rm d}}{{\rm d}t}\begin{pmatrix}
    {a}_{\bm{k}}\\
   {a}^\dagger_{-\bm{k}}
  \end{pmatrix}
  =\begin{pmatrix}
    \epsilon_{{k}}+g(t)\rho&g(t)\rho\\
    -g(t)\rho&-(\epsilon_{{k}}+g(t)\rho)
  \end{pmatrix}\begin{pmatrix}
    {a}_{\bm{k}}\\
   {a}^\dagger_{-\bm{k}}
  \end{pmatrix}.
\end{equation}
Introduce Bogoliubov transformation, $a_{\bm{k}}=u_k(t)b_{\bm{k}}+v_k(t)b_{-\bm{k}}^\dagger$, where
\begin{equation}
  u_k(t)\pm v_k(t)=\left[\frac{\epsilon_k}{\hbar\omega_k(t)}\right]^{\pm 1/2},(\hbar\omega_k(t))^2=\epsilon_k(\epsilon_k+2g(t)\rho).
\end{equation}

Define $\mathcal{A}_{\bm{k}}=(a_{\bm{k}}, a^\dagger_{-\bm{k}})^{\rm T}$, $\mathcal{B}_{\bm{k}}=(b_{\bm{k}}, b^\dagger_{-\bm{k}})^{\rm T}$. The connection of the two operators is
\begin{equation}
 \mathcal{ A}_{\bm{k}}(t)=\mathcal{W}(t,t_0)\mathcal{B}_{\bm{k}}(t_0),\label{AB}
\end{equation}
where evolution matrix $\mathcal{W}(t,t_0)$ is expressed as
\begin{equation}
  \mathcal{W}(t,t_0)=\begin{pmatrix}
    U(t,t_0)&V^*(t,t_0)\\
    V(t,t_0)&U^*(t,t_0)
  \end{pmatrix}.
\end{equation}

To calculate the Bogoliubov weights $U(t,t_0)$ and $V(t,t_0)$, we define the harmonic operators
\begin{align}
  q_{\bm{k}}&=\frac{1}{k}(a_{\bm{k}}+a^\dagger_{-\bm{k}}),\\
  p_{-\bm{k}}&=\frac{\hbar k}{2{\rm i}}(a_{\bm{k}}-a^\dagger_{-\bm{k}}).
\end{align}
Using the harmonic operators to rewrite Hamiltonian (\ref{H}), we find that $q_{\bm{k}},p_{-\bm{k}}$ obey the dynamical equations,
\begin{align}
  \dot{q}_{\bm{k}}&=\frac{p_{-\bm{k}}}{m},\\ \dot{p}_{-\bm{k}}&=-m\omega^2_{k}(t)q_{\bm{k}}.
\end{align}
The solutions to the above equations for given initial conditions $q_{\bm{k}}(t_0)$ and $p_{-\bm{k}}(t_0)$ may be expressed as
\begin{align}
  q_{\bm{k}}(t)&=\gamma_{1}(t,t_0)q_{\bm{k}}(t_0)+\frac{\gamma_2(t,t_0)}{m\epsilon_k/\hbar}p_{-\bm{k}}(t_0)\label{e1},\\
  p_{-\bm{k}}(t)&=m\dot{\gamma}_1(t,t_0)q_{\bm{k}}(t_0)+\frac{\dot{\gamma}_2}{\epsilon_k/\hbar}p_{-\bm{k}}(t_0).\label{e2}
\end{align}
After a few calculations, we find that the Bogoliubov weights and the evolution functions $\gamma_1,\gamma_2$ satisfy the following relations:
\begin{align}
  U(t,t_0)+V(t,t_0)&=\sqrt{\frac{\epsilon_k}{\hbar\omega_k(t_0)}}\gamma_k(t,t_0),\label{uv1}\\
  U(t,t_0)-V(t,t_0)&=\sqrt{\frac{\hbar\omega_k(t_0)}{\epsilon_k}}\gamma'_k(t,t_0),\label{uv2}
\end{align}
where
\begin{equation}
  \begin{aligned}
  \gamma_k(t,t_0)&=\gamma_1(t,t_0)-{\rm i}\frac{\hbar\omega_k(t_0)}{\epsilon_k}\gamma_2(t,t_0),\\ \gamma_k'(t,t_0)&=\frac{{\rm i}\dot{\gamma}_1(t,t_0)}{\omega_k(t_0)}.\label{su}
\end{aligned}
\end{equation}

The evolution function $\gamma_k$ fulfills the equation,
\begin{equation}
  \ddot{\gamma_k}+\omega^2_k(t)\gamma_k=0.\label{efe}
\end{equation}
with initial conditions
\begin{equation}
  \gamma_k(t_0,t_0)=1,\dot{\gamma}_k(t_0,t_0)=-{\rm i}\omega_k(t_0).\label{cond}
\end{equation}
Equation(\ref{efe}) is the key equation to the dynamical theory. The initial condition (\ref{cond}) can be derived from (\ref{e1}), (\ref{e2}), and (\ref{su}).

The expressions of momentum distribution $n_{\bm{k}}(\tau)$ and the dynamical structure factor $S(\bm{k},\tau)$ are
\begin{align}
  n_{\bm{k}}\left(\tau\right)&=|V_k(\tau)|^{2}+\left(|U_k(\tau)|^{2}+|V_k(\tau)|^{2}\right) N_{\bm{k}}\left({0}\right),\label{nk}\\
  S(\bm{k}, \tau)&=\frac{\hbar\omega_k}{\epsilon_{k}}\left|U_k(\tau)+V_k(\tau)\right|^{2} S\left(\bm{k}, {0}\right),\label{sk}
\end{align}
where
\begin{align}
  N_{\bm{k}}(0)&=\langle b_{\bm{k}}^\dagger(0) b_{\bm{k}}(0)\rangle, \\ S(\bm{k},0)&=\frac{\epsilon_k}{\hbar\omega_k}[2N_{\bm{k}}(0)+1].
\end{align}
The Bose gas is initially at a thermal equilibrium state. $N_{{k}}$ is the Bose-Einstein distribution,
  \begin{equation}
    N_{\bm{k}}(0)=\frac{1}{{\rm e}^{{\hbar\omega_k(0)}/{k_BT}}-1}.
  \end{equation}
%%%%========================================================================
\section{The Dynamics with Periodic Modulation}\label{Peri}
%%%%========================================================================
In this section, we will calculate (\ref{efe}) in the case of the periodic modulation and show the evolution of the momentum distribution and dynamical structure factor.

Let $g(t)=g_0\cos \Omega t$, where $g_0$ is a time-independent constant. Equation(\ref{efe}) becomes
\begin{equation}
 \frac{{\rm d}^2\gamma_k}{{\rm d}\tau^2}+(\lambda-2q\cos2\tau)\gamma_k=0\label{me},
\end{equation}
where
\begin{equation}
\tau=\Omega t/2,\quad \lambda=\frac{4\epsilon_k^2}{\Omega^2\hbar^2},\quad q=-\frac{4\rho g_0\epsilon_k}{\Omega^2\hbar^2}.\label{para}
\end{equation}
Equation (\ref{me}) is known as the canonical form of the Mathieu equation. Mathieu equations have periodic, stable and unstable solutions. The condition of periodic solutions is that the parameters $\lambda$ and $q$ satisfy a series of characteristic relations~\cite{Abramowitz1948, McLachlan1951, Wang1989}. Stable and unstable regions separated by these characteristic relations are shown in Fig. 1.

\begin{figure}[ht]
\centering
\includegraphics[width=0.48\textwidth]{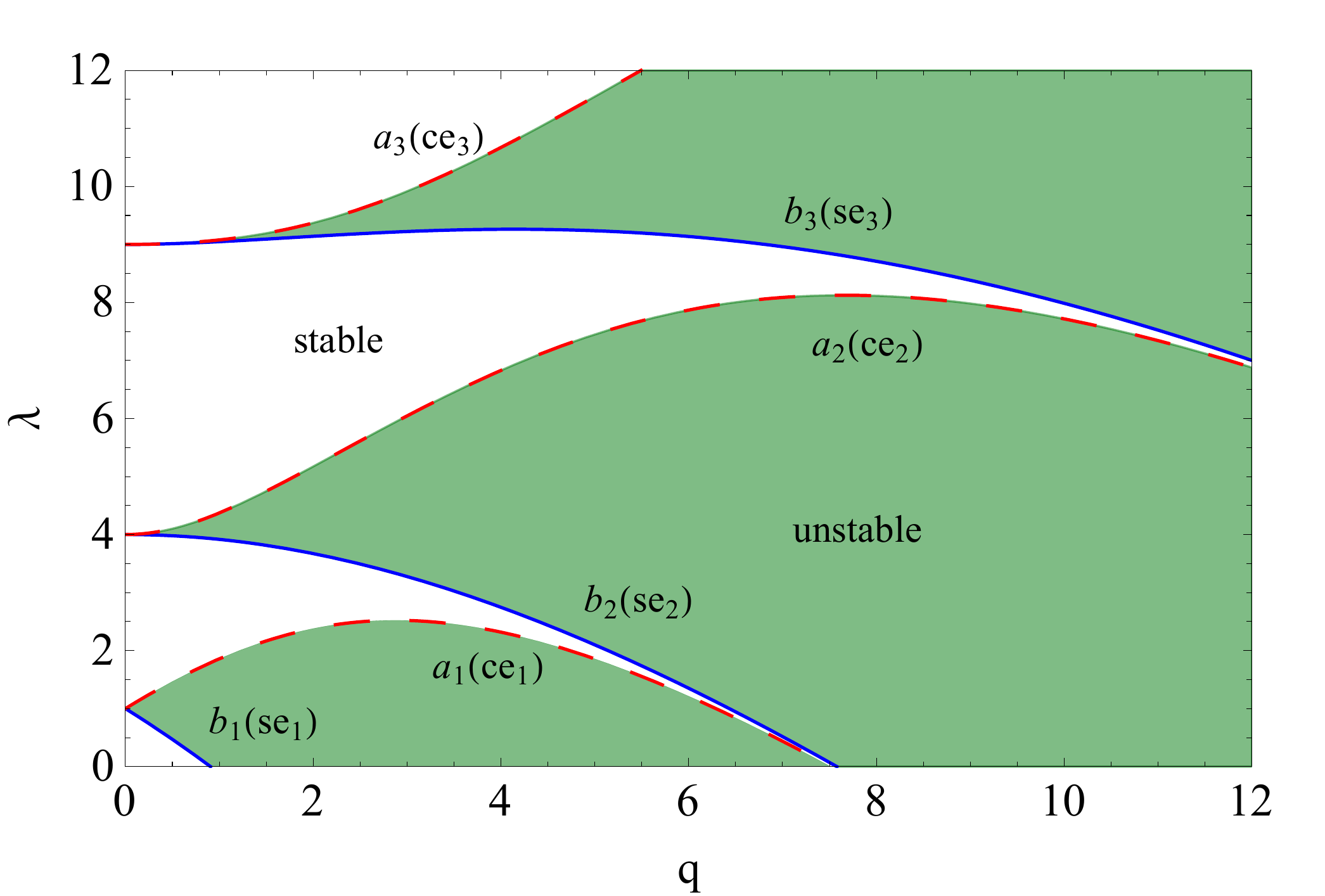}
\caption{\footnotesize Stable and unstable diagram of the Mathieu equation. Each $a_i$ ($i=0,1,2,...$) is an eigenvalue of even periodic solutions ${\rm ce}_i$ of the Mathieu equation (dashed line). Any curve of eigenvalue $b_j$($j=1,2,...$) corresponds to the odd periodical solution ${\rm se}_j$ (solid line). The stable and unstable solution of the Mathieu equation corresponds to the blank areas and shaded areas, respectively.}
\end{figure}

Suppose $y_{1k}(\tau)$ and $y_{2k}(\tau)$ are two linearly independent solutions to (\ref{me}) with the condition when $\tau=0$,
\begin{equation}
\begin{aligned}
  y_{1k}(0)&=1,\dot{y}_{1k}(0)=0;\\
   y_{2k}(0)&=0,\dot{y}_{2k}(0)=1.\label{ic}
\end{aligned}
\end{equation}
The $\dot{y}_{1k}(\tau)$ and $\dot{y}_{2k}(\tau)$ are derivatives with respect to $\tau$. One could demonstrate that $y_{1k}(\tau)$ is an even function and $y_{2k}(\tau)$ is odd. From (\ref{me}) and (\ref{ic}), we obtain a relation
\begin{equation}
  y_{1k}(\tau)\dot{y}_{2k}(\tau)-\dot{y}_{1k}(\tau)y_{2k}(\tau)=1.\label{c1}
\end{equation}

The general solution to (\ref{me}) can be expressed as
\begin{equation}
  \gamma_k(\tau)=A y_{1k}(\tau)+By_{2k}(\tau),
\end{equation}
where $A$ and $B$ are unknown constants. We determine $A=1$, $B=-2{\rm i}\omega_k/\Omega$ with $\omega_k=\omega_k(0)=\sqrt{\epsilon_k^2+2g_0\rho\epsilon_k}/\hbar$ by the initial condition (\ref{cond}). So the expressions of $\gamma_k(\tau)$ and $\gamma_k'(\tau)$ are
\begin{align}
  \gamma_k(\tau)&=y_{1k}(\tau)-{\rm i}\frac{2\omega_k}{\Omega} y_{2k}(\tau),\\
  \gamma_k'(\tau)&=\dot{y}_{2k}(\tau)+{\rm i}\frac{\Omega}{2\omega_k}\dot{y}_{1k}(\tau),
\end{align}
respectively. We derive the expressions of $U(\tau)$ and $V(\tau)$ from (\ref{uv1}) and (\ref{uv2}),
\begin{equation}
  \begin{aligned}
    U_k(\tau)=\frac{1}{2}\sqrt{\frac{\epsilon_k}{\hbar\omega_k}}\left[\left(y_{1k}(\tau)+\frac{\hbar\omega_k}{\epsilon_k}\dot{y}_{2k}(\tau)\right)\right.\\
    \left.-{\rm i}\left(\frac{2\omega_k}{\Omega}y_{2k}(\tau)-\frac{\hbar\Omega}{2\epsilon_k}\dot{y}_{1k}(\tau)\right)\right],\\
     V_k(\tau)=\frac{1}{2}\sqrt{\frac{\epsilon_k}{\hbar\omega_k}}\left[\left(y_{1k}(\tau)-\frac{\hbar\omega_k}{\epsilon_k}\dot{y}_{2k}(\tau)\right)\right.\\
     \left.-{\rm i}\left(\frac{2\omega_k}{\Omega}y_{2k}(\tau)+\frac{\hbar\Omega}{2\epsilon_k}\dot{y}_{1k}(\tau)\right)\right].
  \end{aligned}\label{uv}
\end{equation}

Substituting (\ref{uv}) into (\ref{nk}) and (\ref{sk}), and considering condition (\ref{c1}) we obtain
\begin{align}
   n_{\bm{k}}(\tau)=\frac{1}{4}\frac{\epsilon_k}{\hbar\omega_k}\left[1+2N_{\bm{k}}(0)\right]\left[n_{1k}(\tau)+\frac{4\omega^2_k}{\Omega^2}n_{2k}(\tau)\right]-\frac{1}{2},\label{nfc}
   \end{align}
   \begin{align}
S(\bm{k},\tau)=\left[y_{1k}^2(\tau)+\frac{4\omega_k^2}{\Omega^2}y_{2k}^2(\tau)\right]S(\bm{k},0),\label{sfc}
\end{align}
where
\begin{equation}
  \begin{aligned}
    n_{1k}(\tau)&=y_{1k}^2(\tau)+\frac{\hbar^2\Omega^2}{4\epsilon^2_k}\dot{y}^2_{1k}(\tau),\\
    n_{2k}(\tau)&=y_{2k}^2(\tau)+\frac{\hbar^2\Omega^2}{4\epsilon^2_k}\dot{y}^2_{2k}(\tau).
  \end{aligned}\label{n1n2}
\end{equation}
Equations (\ref{nkp}) and (\ref{skp}) indicate that the shapes of evolution curves of $n_k(\tau)$ and $S(k,\tau)$ are not affected by the detailed expressions of $N_{\bm{k}}(0)$ and $S(\bm{k},0)$. Therefore, we just need to consider the case of zero temperature at the initial time. Considering $N_{\bm{k}}(0)=0$ and $S(k,0)=\epsilon_k/\hbar\omega_k$, we achieve
\begin{equation}
  n_{\bm{k}}(\tau)=\frac{1}{4}\frac{\epsilon_k}{\hbar\omega_k}\left[n_{1k}(\tau)+\frac{4\omega_k^2}{\Omega^2}n_{2k}(\tau)\right]-\frac{1}{2},\label{nkp}
\end{equation}
\begin{equation}
  S(\bm{k},\tau)=y_{1k}^2(\tau)+\frac{4\omega_k^2}{\Omega^2}y_{2k}^2(\tau).\label{skp}
\end{equation}

Introducing $\varepsilon_0=\hbar\Omega/2$ and considering $\lambda$ and $q$ in (\ref{para}), we can express equations (\ref{nkp}) and (\ref{skp}) as
\begin{equation}
  n_{\bm{k}}(\tau)=\frac{1}{4}\sqrt{\frac{\lambda}{\lambda-2q}}\left[n_{1k}(\tau)+(\lambda-2q)n_{2k}(\tau)\right]-\frac{1}{2},\label{nkp2}
  \end{equation}
  \begin{equation}
  S(\bm{k},\tau)=y_{1k}^2(\tau)+(\lambda-2q)y_{2k}^2(\tau),\label{skp2}
\end{equation}
with
\begin{equation}
  \begin{aligned}
 n_{1k}(\tau)&=y_{1k}^2(\tau)+\frac{1}{\lambda}\dot{y}^2_{1k}(\tau),\\
    n_{2k}(\tau)&=y_{2k}^2(\tau)+\frac{1}{\lambda}\dot{y}^2_{2k}(\tau).
\end{aligned}
\end{equation}

Comparing the derivatives of $n_{\bm{k}}(\tau)$ and $S(\bm{k},\tau)$ respect to time in (\ref{nkp2}) and (\ref{skp2}), and considering Mathieu equation (\ref{me}), we derive a concise relation
\begin{equation}
  \dot{n}_{\bm{k}}(\tau)=\frac{q\cos2\tau}{2\sqrt{\lambda(\lambda-2q)}}  \dot{S}(\bm{k},\tau) .\label{diff}
\end{equation}
This is the main result of the present work. This relation implies that the extreme values of $n_{\bm{k}}(\tau)$ and $S(\bm{k},\tau)$ emerge at the same time except $\tau=(2n+1)\pi/4, n=0,1,2...$ The $\cos2\tau$ in (\ref{diff}) originates from the periodical modulation $g=g_0\cos2\tau$.

In general many-body systems, there is no direct relationship between the properties of the single-particle property and two-body correlation. However, in Bose-Einstein condensates, they are strongly related to each other~\cite{Watabe2020}. Our result indicates that, for evolutive Bose-Einstein condensates, single-particle property (momentum distribution) strongly relates to two-body correlation (dynamical structure factor). This phenomenon will be shown in the next section.

\section{evolution curves of $n_{\bm{k}}(\tau)$ and $S(\bm{k},\tau)$ }\label{ens}
The Mathieu equation has a property that for each characteristic value, there is only one periodical solution. Therefore, ${y}_{1k}(\tau)$ and ${y}_{2k}(\tau)$ can not be periodic solutions simultaneously. In other words, $n_{\bm{k}}(\tau)$ and $S(\bm{k},\tau)$ have no strictly periodic behavior. Thus we will discuss the stable and unstable dynamics in this section.

\subsection{Stable dynamics }

We consider the stable dynamics by taking parameters $\lambda=2$ and $q=0.1$ in the blank region of Fig.1. By numerically solving the Mathieu equation with the initial condition (\ref{ic}), we obtain $y_{1k}(\tau)$ and $y_{2k}(\tau)$. Then we calculate the momentum distribution and dynamical structure factor with the consideration of (\ref{nkp2}) and (\ref{skp2}). The results are shown in Fig.2.
\begin{figure}[ht]
\centering
\includegraphics[width=0.48\textwidth]{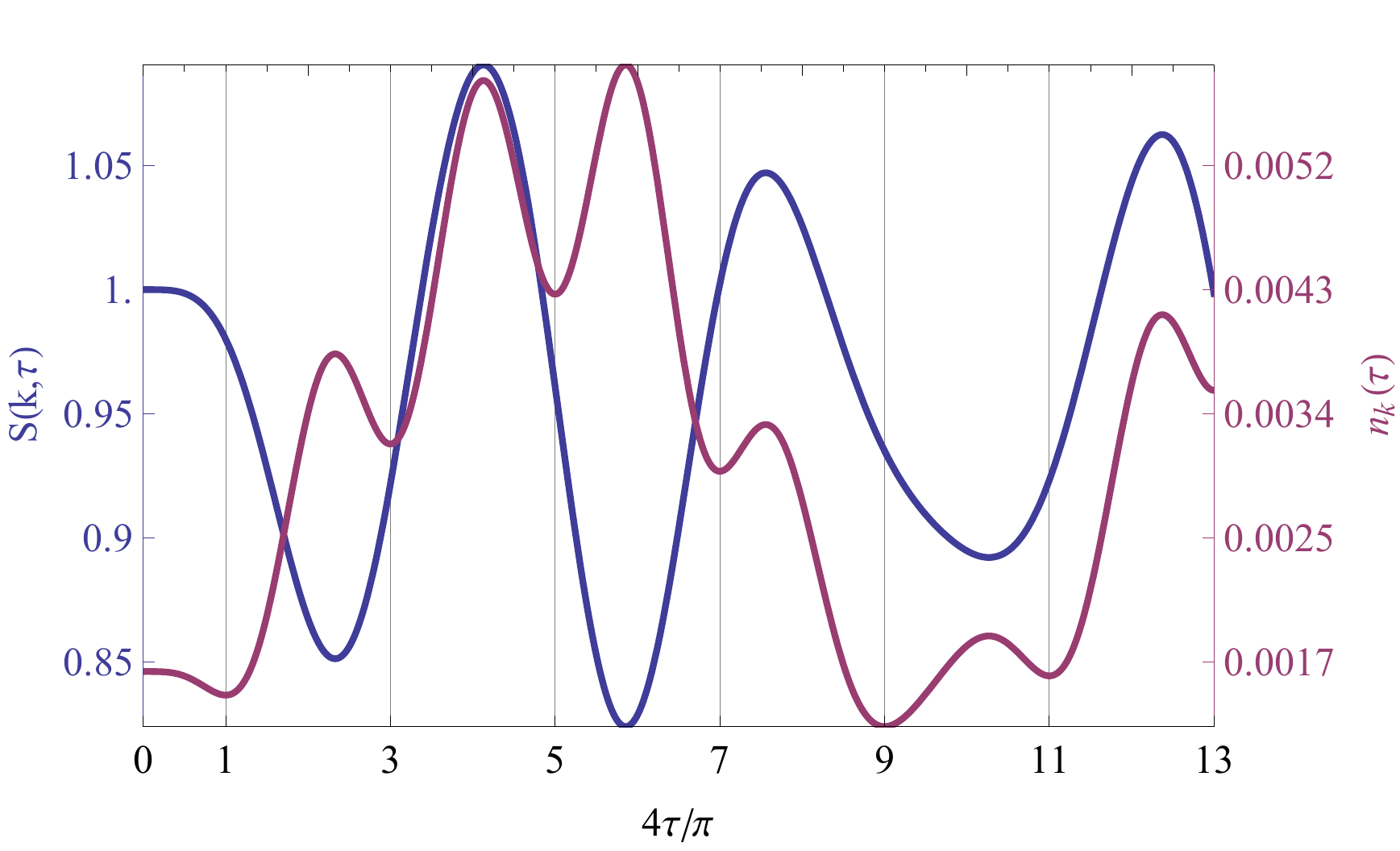}
\caption{The stable dynamics for $\lambda=2,q=0.1$. The peaks of momentum distribution and dynamical structure factor synchronously emerge except $\tau=(2n+1)\pi/4,n=0,1,2,...$, which shows a strong relation of one-body property and two-body correlation. }
\end{figure}

We observe that some peaks in the curves of the momentum distribution and structure factor appear at the same time. In other words, these peaks are synchronous. In particular, these synchronous peaks are not located in $\tau=(2n+1)\pi/4$, which is consistent with (\ref{diff}).

\subsection{Unstable dynamics}
We consider the stable dynamics by taking parameters $\lambda=1$ and $q=0.1$ in the shadowed region of Fig.1. By numerically solving the Mathieu equation with the initial condition (\ref{ic}), we obtain $y_{1k}(\tau)$ and $y_{2k}(\tau)$. Then we calculate the momentum distribution and dynamical structure factor with the consideration of (\ref{nkp2}) and (\ref{skp2}). The results are shown in Fig.3.

Even in a weak-interacting Bose gas, as the system evolves with time, the momentum distribution and structure factor become larger and larger. It can be analogous to the Faraday wave that emerged in the real space of Bose-Einstein condensation~\cite{Engels2007, Kagan2001, Staliunas2011, Nicolin2011}. The peaks are growing in exponential form because of the emergence of parametric resonances~\cite{Atas2019}. Such exponential growth can also be found in time-dependent Bogoliubov theory in~\cite{Wu2019}.
\begin{figure}[ht]
\centering
\includegraphics[width=0.48\textwidth]{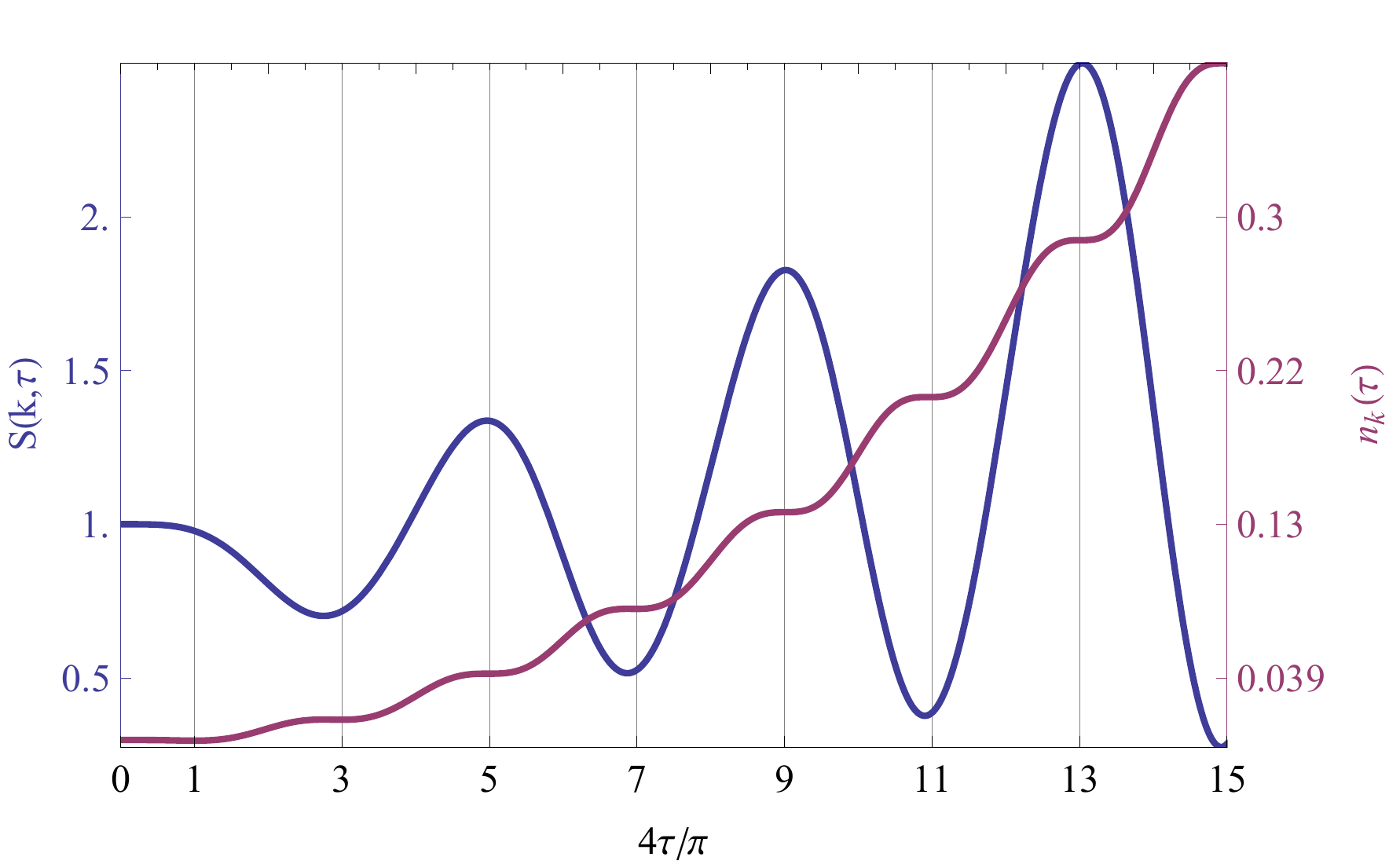}
\caption{\footnotesize  The unstable dynamics for $\lambda=1,q=0.1$. The evolving dynamical structure factor and momentum distribution are all exponentially increasing. The evolving dynamical structure factor has evident peak structure, while the momentum distribution becomes flat at $\tau=(2n+1)\pi/4$.}
\end{figure}

Unlike the case of stable dynamics, although the dynamical structure factor has evident peak structure, however, the momentum distribution has no such evident peaks. This makes us observe synchronous evolutions difficultly.

\section{conclusion}\label{conc}
In this paper, we analytically show that the time evolution of momentum distribution $n_{\bm{k}}(t)$ and dynamical structure factor $S(\bm{k},t)$ in a periodical modulation of weak Bose gas.  With the cosine-modulated interaction of the system, the evolution equations become Mathieu equations that have three-type solutions, namely three-type dynamics. However, we note that it is impossible to emerge periodical dynamics of the momentum distribution and the dynamical structure factor.

We derive an exact relation between the derivatives of the momentum distribution and the dynamical structure factor in periodic modulation. We show that the stable dynamics with the amplitudes of $n_{\bm{k}}(t)$ and $S(\bm{k},t)$ remain finite. It is found that some peaks of the momentum distribution and the dynamical structure factor are synchronous, which is an agreement with the derivative relation. This relation indicates that momentum distribution (single-particle property) strongly related to the dynamical structure factor (two-body correlation) in evolving Bose-Einstein condensates. The unstable dynamics of Bose gas, similar to the instability of classical liquid interface, can also emerge the Faraday waves~\cite{Nguyen2019}. We find the unstable dynamics which have no evident synchronous peak structure of momentum distribution and dynamical structure factor. The increasing peaks possess exponential behavior. Very recently, Cheng and Shi have found that the SU(1,1) group leads to this exponential behavior~\cite{Cheng2020}. We believe that the results in our work will be confirmed in experiments of ultracold gas.

\section*{Acknowledgement}

The authors are grateful for financial support from the National Natural Science Foundation of China (Grants No. 11675017 and No. 11975050). We are very grateful for the discussion with Yiheng Zhang.

\end{document}